\documentclass[twocolumn,showpacs,preprintnumbers,amsmath,amssymb]{revtex4}

\usepackage{graphicx}
\usepackage{dcolumn}
\usepackage{bm}

\begin{document}

\preprint{}

\title{Does the interaction potential determine both the fragility of a
liquid and the vibrational properties of its glassy state~?}

\author{Patrice Bordat, Fr\'ed\'eric Affouard, Marc Descamps}
 \email{bordat@cyano.univ-lille1.fr}
 \affiliation{
Laboratoire de Dynamique et Structure des Mat\'eriaux Mol\'eculaires\\
UMR 8024, Universit\'e Lille I, 59655 Villeneuve d'Ascq cedex, France
}
\author{K. L. Ngai}
\affiliation{
Naval Research Laboratory, Washington, DC 20375-5320, USA
}

\date{\today}

\begin{abstract}
By performing molecular dynamics simulations of binary Lennard-Jones systems
with three different potentials, we show that increase of anharmonicity and
capacity for intermolecular coupling of the potential is the cause of (i) the
increase of kinetic fragility and nonexponentiality in the liquid state, (ii)
the $T_{g}$-scaled temperature dependence of the nonergodicity parameter
determined by the vibrations at low temperatures in the glassy state.
Naturally these parameters correlate with each other, as observed
experimentally by T. Scopigno {\it et al.}, Science
\textbf{302}, 849 (2003).
\end{abstract}


\maketitle

The structural relaxation time, $\tau $, of all glass-forming liquids increases on 
cooling. It becomes so long at some temperature $T_{g}$ that equilibrium 
cannot be maintained and the liquid is transformed to a glass. $T_{g}$ is 
defined as the temperature at which $\tau $ reaches some arbitrarily chosen long 
time, say 10$^{2}$ s. Although this behavior is shared by glass-formers of 
diverse chemical and physical structures,
the scaled temperature dependence of $\tau $ in the 
liquid state can differ greatly from one liquid to another in the degree of 
departure from the Arrhenius scaled temperature dependence~\cite{Laughlin1972,Angell1991}.
The departure can be characterized 
by the rapidity of the change of log$\left( \tau \right)$ with $T_{g}$/$T$ at $T_{g}$/$T$=1, which is 
given by the steepness index or the fragility $m$ defined by~\cite{PlaBoh9192}
\begin{equation}
\label{eq1}
m=\left. \frac{d\log (\tau)}{d(T_g /T)} \right|_{T_g /T=1} .
\end{equation}
The values of $m$ of glass-formers of all kinds vary over a large range, from 
the least value of about 17 for strong glass-formers (like silica) 
having nearly Arrhenius scaled temperature dependence of $\tau $, to values as high 
as about 200 found for some glass-formers called fragile. Naturally, such 
large variations observed in $m$ beg the question of its microscopic origin. 
Several attempts have been made in the past to correlate $m$ with other dynamic 
or thermodynamic properties, with the hope that the correlations will lead 
to the factor or factors that determine $m$. Examples include: (1) The 
correlation of $m$ with $n \equiv \left( 1-\beta \right)$ at $T=T_{g}$,
where $\beta$ is the stretched exponent in the 
Kohlrausch function, $\exp[- \left(t/\tau \right)^{\beta}]$, used to fit the time dependence of 
the correlation functions such as the intermediate scattering functions.
(2) The correlation of $m$ or $\left( 1-\beta \right)$ with 
the mean square displacement $<u^{2}>$ obtained \cite{Ngai2000}
from quasielastic neutron 
scattering measurement of the Debye-Waller factor
$\exp[-<u^{2}>Q^{2}/3]$. Glass-former with larger $m$ or 
$\left( 1-\beta \right)$ has a larger $<u^{2}>$ at the same value of $T/T_{g}$ and rise more 
rapidly as a function of $T/T_{g}$, below $T_{g}$ as well as near and across 
$T_{g}$ in the liquid states~\cite{Ngai2000}.
(3) The correlation between $m$ and the slope of the change of the 
configurational entropy, $S_{c}$, with $T/T_{g}$ at $T_{g}$~\cite{Ito1999}.
(4) The correlation 
of $m$ with the statistics of potential energy minima of the energy landscape~\cite{Speedy1999,Sastry2001}.
(5) The correlation of $m$ with the temperature dependence of the 
shear modulus of the liquid~\cite{Dyre}.
Perhaps the most intriguing of all correlations is (6) between $m$ and the 
vibrational properties of the glass at temperatures well below $T_{g}$ found 
recently by T.~Scopigno {\it et al.}~\cite{Scopigno2003}.
The nonergodicity parameter, $f \left( Q,T \right)$, at $Q$=2 nm$^{-1}$ 
at very low temperatures is determined by vibrations. From inelastic X-ray 
scattering data, its temperature dependence is well described by $[1+\alpha \left( T/T_{g} \right)]^{-1}$. 
T.~Scopigno showed that $m$ and $\alpha$ are proportional for many glass-formers. Apparently, 
this last correlation (6) seems to be related to (2) for 
$<u^{2}(T/T_{g})>$ from neutron scattering at temperatures well below 
$T_{g}$.

In any glass-former, it is the interaction potential, $V(r)$, that determines 
ultimately all dynamic, thermodynamic and vibrational properties at all 
temperatures, below and above $T_{g}$. Changes in any of quantities, $m$, $n$, 
$<u^{2}(T/T_{g})>$, $\alpha $, $S_{c}$, free volume $\nu$~\cite{Ngai2001}
and the degree 
of dynamic heterogeneity, from one glass-former to another originate from 
the change in $V(r)$. Thus, correlations found between these quantities are 
clues for finding out which aspects of $V(r)$ determine them and give rise to 
the correlations between them. One would like to examine the interaction 
potentials in real glass-formers. However, in such materials, the different kinds of chemical 
bonding and the different sizes of the basic structural unit make the 
comparisons ambiguous. For this reason we consider the 
binary Lennard-Jones particles with different choices of interaction 
potentials $V(r)$ between the particles, and perform molecular dynamics (MD)
simulations on them to obtain $m$, $n$, $\alpha$, and $<u^{2}(T/T_{g})>$. Correlations 
are found between all these quantities, thus reproducing the empirical 
findings from real glass-formers. Since the number of particles as well as 
their density are the same, the changes of these quantities are 
predominantly due to the change in $V(r)$. The latter is well controlled, and therefore
we identify anharmonicity and the capacity of intermolecular coupling 
of $V(r)$ to be responsible for enhancement of $m$, $n$, $\alpha$, and 
$<u^{2}(T/T_{g})>$, and hence their correlations.

\begin{figure}[htbp]
\includegraphics[scale=0.33,clip=true,angle=-90]{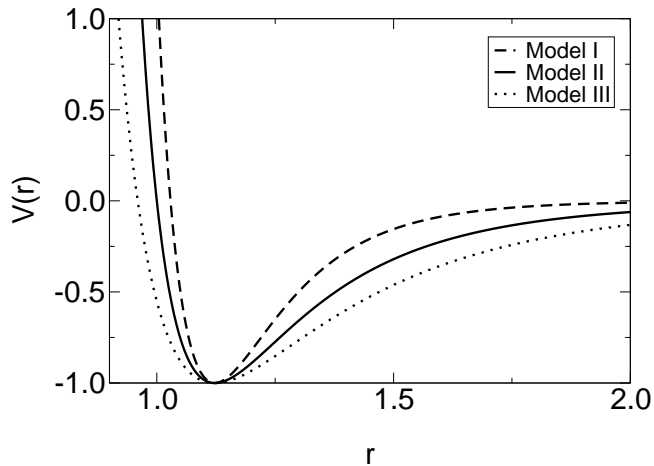}
\caption{\label{fig1} Potential $V$(r) governing the A-A interaction. The dashed curve is
the (8,5) LJ potential for model I, the solid curve is the (12,6) LJ
potential for model II and the dotted curve is the (12,11) LJ potential for
model III.}
\end{figure}
MD simulations were performed on binary Lennard-Jones (LJ) 
particles systems with three different interaction potentials by the 
MD package DLPOLY~\cite{DLPOLY}.
Technical details of the MD simulations are given in \cite{Bordat2003}. We have
performed from $10^5$ to $6 \times 10^7$ time steps, depending on temperature. We have investigated
12 temperatures in the range of $[0.675 - 5]$, $[0.416 - 5]$ and $[0.26 - 2]$ for Model I, II and III
respectively. All 
models are composed of 1500 uncharged particles (1200 species A and 300 
species B). The generalized $(q,p)$ LJ potentials have the form,
$V(r) = \frac{E_{0}}{(q-p)}\left(p\left(\frac{r_{0}}{r}\right)^{q} - q\left(\frac{r_{0}}{r}\right)^{p}\right)$. 
The parameters $r_{0}$ and $E_{0}$ represent the position of the minimum of the well and its 
depth, respectively. The reduced LJ units~\cite{Allen1987} are used.
The choice of $q$=12 and $p$=6 corresponds to 
the standard LJ potential used by Kob {\&} Andersen (K-A)~\cite{Kob9597} and others
for extensive studies by simulation. For the purpose of investigating the 
change of dynamics with controlled change of $V(r)$, we developed two other 
models by changing only the exponents, $q$ and $p$, of the LJ potential for the 
A - A interactions. They are ($q$=8, $p$=5) and ($q$=12, $p$=11) and 
shown together with the (12,6) LJ potential in Fig.~\ref{fig1}. The well depth and 
the position of the minimum of $V(r)$ are unchanged, and we have kept the 
standard (12,6) LJ potentials of the K-A model for the A - B and B 
- B interactions, in order to retain as much as possible the remarkable 
ability of the K-A model to form a glass upon cooling. The (12,11) 
LJ potential is more harmonic than the classical (12,6) LJ potential, while 
the (8,5) LJ potential is a flat well and exceedingly anharmonic. 
Whenceforth the (12,11), (12,6) and (8,5) potentials are referred to as 
Model I, II and III respectively, reminding us that anharmoncity is 
increasing in this order. 

The structure has been briefly studied by the radial distribution functions $g(r)$ of
the (12,11) and (8,5) models, which have 
been calculated and found very similar to that of (12,6) model in the 
temperature ranges investigated. Respectively for models I, II and III, the 
position of the first peak of $g(r)$ is at 1.072, 1.066 and 1.057 and its width 
at half the maximum is 0.129, 0.147 and 0.191. The $g(r)$ of the three models in 
the supercooled state is similar to that in the liquid regime, ensuring the 
structures have disorder in the mid- and long-range.

\begin{figure}[htbp]
\includegraphics[scale=0.33,clip=true,angle=-90]{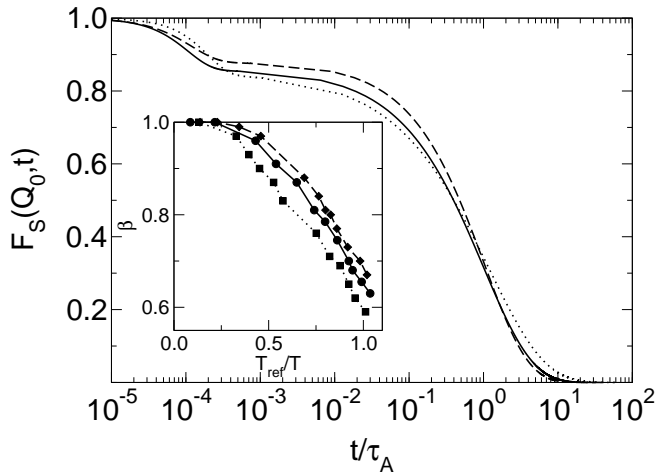}
\caption{\label{fig2} Self intermediate scattering function $F_{S}(Q_0,t)$ versus scaled time
$t/\tau_{A}$. Dashed, continuous and dotted lines are for Models I, II and III 
respectively. For all three models, $\tau_{A}(T_{ref})$=~46435.8. 
The inset shows the stretched exponent $\beta=(1-n)$ as a function of the 
scaled reciprocal temperature $T_{ref}/T$ for the three models. ($\blacklozenge$)
Model I, ({\Large $\bullet$}) Model II and ($\blacksquare$) Model III. For definitions of 
$\tau_{A}$ and $\beta$, see text.}
\end{figure}

Dynamics have been 
investigated by computing the self $F_{S}(Q,t)$ and the total $F(Q,t)$ intermediate
scattering functions of particles A at $Q_0=2 \pi/r_{0}$ for the three models. At high 
temperatures, $F_{S}(Q_0,t)$ decays linear exponentially to zero with a 
characteristic time of about 0.45 close to crossover time $t_{c}\approx$ 1 to 2~ps
used as a fundamental time in the Coupling Model (CM)~\cite{Ngai1999b,Ngai2003}.
When temperature is lowered, the dynamics slows down 
dramatically and a two-step process appears. This behavior is well described
by the Mode Coupling Theory (MCT)~\cite{Gotze1992}.
From the 
$F_{S}(Q_0,t)$ at each of these lower temperatures, we determine the nonergodicity
parameter (height of the plateau), $f_S(Q_0,T)$, the relaxation time, $\tau_{A}$, and the stretched 
exponent, $\beta$, from the fit to the second step decay of $F_{S}(Q_0,t)$ by 
$f_S(Q_0,T)\exp[-(t/\tau_{A})^{\beta}]$. Shown in Fig.~\ref{fig2} are $F_{S}(Q_0,t)$ versus $t/\tau_{A}$ 
of all three models at the reference temperature $T_{ref}$ defined 
by $\tau_{A}(T_{ref})$=~46435.8, a very long time in our simulations. 
The values of $T_{ref}$ are 0.688, 0.431 and 0.263 for Models I, II
and III respectively.

\begin{figure}[htbp]
\includegraphics[scale=0.33,clip=true,angle=-90]{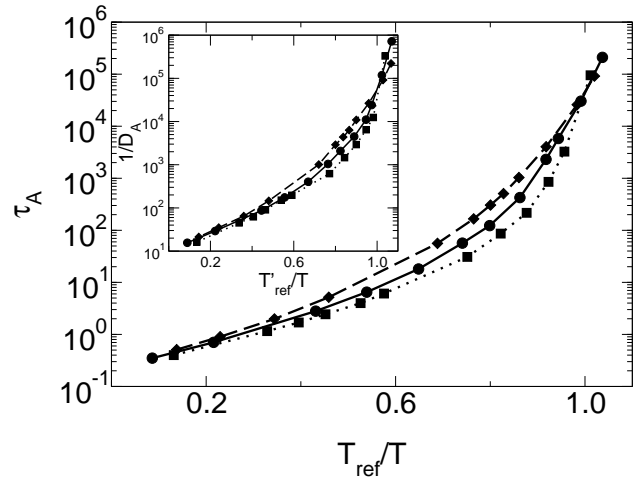}
\caption{\label{fig3} The relaxation times $\tau_{A}$ obtained from $F_{S}(Q_0,t)$ for the
three models as a function of $T_{ref}/T$ where $T_{ref}$ is defined as the temperature at which
$\tau_{A}$ reaches 46435.8. ($\blacklozenge$)
Model I, ({\Large $\bullet$}) Model II and ($\blacksquare$) Model III.
$T_{ref}$ is the analogue of $T_g$
for simulations when the dynamics of the system slows down to more than
10~ns. In the inset, the reciprocal of diffusivity $D_{A}$ of A particles
for the three models are given as a function of $T'_{ref}/T$. Here
$T'_{ref}$ is now defined as the temperature at which $D_{A}$ is equal to $1.857432 \times 10^{-5}$.}
\end{figure}

Shown in Fig.~\ref{fig2} (inset) are $\beta$ of the three models as a function of $T_{ref}/T$.
At any $T_{ref}/T$, $(1-\beta)$ is least for model I 
and largest for model III. At $T=T_{ref}$, $\beta $=0.69, 0.65 and 0.60 respectively 
for Models I, II and III. Thus $(1-\beta)$ increases with 
anharmonicity. In Fig.~\ref{fig3}, $\log(\tau_{A})$ is plotted against $T_{ref}/T$ and the data 
of the three models show systematic change. It can be seen that the slope, fragility index
$m(\tau_{A}) \equiv \frac{d\log(\tau_{A})}{d(T_{ref}/T)}$ as $(T_{ref}/T)\to 1$, increases
monotonically in the order of Models I, II and III. The values of $m(\tau_{A})$ determined
from this latter relation are 15.07, 18.57 and 26.58 for Models I, II and III.
Alternatively, the estimated values of 
$m(\tau_{A})$ based on Sastry's method~\cite{Sastry2001} are 0.195, 0.241 and 0.405
for Models I, II and III, respectively. 
Hence, $m(\tau_{A})$ increases with anharmonicity. The fragility index found in 
the present study for model II is in good agreement with the value 
determined in an earlier work on the same model~\cite{Sastry2001}.
The diffusion coefficient, $D_{A}$, of particles A was calculated 
from the mean-square displacement $<u^{2}(t)>$
at long times when 
$<u^{2}(t)>$ assumes the linear $t$ dependence. Here another $T'_{ref}$ is 
defined as the temperature at which $D_{A}$ is equal to $1.86\times 
10^{-5}$. In the inset of Fig.~\ref{fig3}, we plot $\log(1/D_{A})$ against 
$T'_{ref}/T$ for each of the three models. They exhibit the same pattern as 
$\log(\tau_{A})$. Again the steepness or fragility index, $m(D_{A}) \equiv
\frac{d \log(1/D_{A})}{d(T'_{ref}/T)}$ as $(T'_{ref}/T)\to 1$, increases monotonically in 
the order of Models I, II and III, or with anharmonicity.

\begin{figure}[htbp]
\includegraphics[scale=0.33,clip=true,angle=-90]{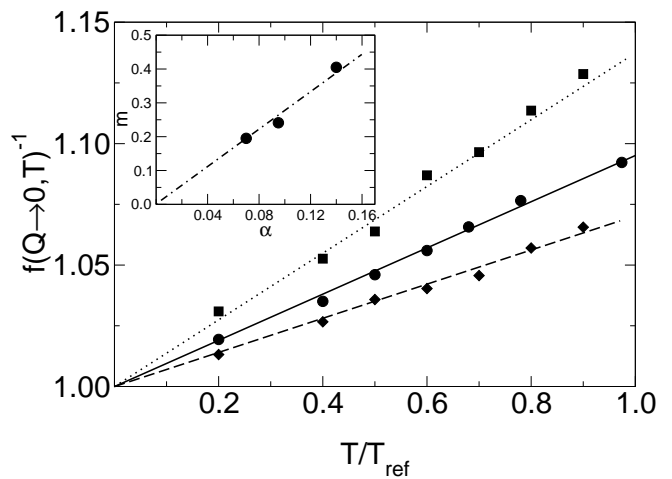}
\caption{\label{fig4} $f(Q,T)^{-1}$ versus $T/T_{ref}$ for the three models.
($\blacklozenge$)
Model I, ({\Large $\bullet$}) Model II and ($\blacksquare$) Model III.
$f(Q,T)^{-1}$ is almost linear
relative to $T/T_{ref}$ with a slope noted $\alpha$. The inset shows the
correlation of the fragility $m$ with $\alpha$ from the results of the three
models.}
\end{figure}

So far we are concerned for dynamic quantities and their correlations for $T > T_{ref}$,
as analogues of them in the liquid state of real glass-formers. 
Next we examine the vibrational properties of the three models at 
temperatures lower and much lower than $T_{ref}$. At low temperatures, 
relaxation of any kind is absent in the simulation time window, and the 
nonergodicity parameter $f(Q,T)$ determined from $F(Q,t)$ is contributed 
entirely from vibrations. As performed in~\cite{Scopigno2003}, we have followed
the behaviour of $f(Q \to 0,T)$ which has been obtained by a Q-quadratic extrapolation
of $f(Q,T)$ for the lowest temperatures. The results of the three
models are shown by a plot of $f(Q \to 0,T)^{-1}$ versus $T/T_{ref}$ in Fig.~\ref{fig4}.

In all three cases, the dependence of $f(Q,T)^{-1}$ on $T/T_{ref}$ is approximately 
linear, and $f(Q,T)^{-1}$ has the extrapolated value of unity at the origin. The 
dependence of $f(Q,T)$ from simulation on $T/T_{ref}$ is governed by the parameter, 
$\alpha$, through the expression,
\begin{equation}
\label{eq2}
f(Q,T)^{-1}=1+\alpha \frac{T}{T_{ref}},
\end{equation}
just like a similar expression used to represent the dependence of 
$f(Q,T)$ on $T/T_{g}$ of real glass-formers obtained by inelastic X-ray scattering~\cite{Scopigno2003}.
We see from Fig.~\ref{fig4} that the increase of 
$f(Q,T)^{-1}$ with $T/T_{ref}$ is fastest for model III and slowest for model I. 
Equivalently stated, the slope $\alpha$ is largest for model III and smallest for 
model I. $\alpha$ increases with anharmonicity of the potential, like $m(\tau_{A})$ or 
$m(D_{A})$, and $(1-\beta)$, as seen before (Figs.~\ref{fig1}, \ref{fig2}, \ref{fig3}). Hence the interacting 
potential is the origin of the correlation of $\alpha$ with $m(\tau_{A})$ or 
$m(D_{A})$, and $(1-\beta)$ in our simulation, suggesting the same 
holds for real glass-formers. Moreover, in the inset of Fig.~\ref{fig4}, we observe that $m(\tau_A)$
and $\alpha$ are proportional together as shown in~\cite{Scopigno2003}.

Presumably there is no disagreement that the interaction potential is 
pivotal in determining all dynamic, thermodynamic and vibrational properties 
of glass-formers at all temperatures. With all these experimentally 
accessible properties originating from the interaction potential, it is not 
surprising to find correlations or anticorrelations between them.
Making this point is one of the motivations of the 
work in demonstrating that the various correlations between $\alpha$,
$m$, $(1-\beta)$ and $<u^{2}(T$/$T_{g})>$ observed in 
real glass-formers are reproduced by simulations as the analogues of 
correlations between $\alpha$, $m \left( \tau_A \right)$, $m \left( D_A \right)$ and $(1-\beta)$
by varying the interaction potential $V(r)$. The results confer a bonus in 
identifying which feature of the interaction potential is responsible for 
enhancement of $\alpha$, $m \left( \tau_A \right)$, $m \left( D_A \right)$ and $(1-\beta)$.
Certainly the anharmonicity of $V(r)$ increases when going from Model I to II 
and III, but one also can observe from Fig.~\ref{fig1} that the energy barrier becomes 
lower and flatter. The latter trend means that neighboring LJ particles are 
more coupled in their motions. The increase in interparticle coupling
from Model I to Model III is consistent with the position of the first peak 
of radial distribution functions $g(r)$ (1.072, 1.066 and 1.057 for Models I, II 
and III respectively) and with its width at half the maximum (0.129, 0.147 
and 0.191 for Models I, II and III). This insight from 
simulation, when transferred to real glass-formers, suggests that the 
capacity for intermolecular coupling and anharmonicity of the interaction 
potential determine the dynamic, thermodynamic and vibrational properties of 
glass-formers above as well as below $T_{g}$. Although the thermodynamic 
variables, configurational entropy $S_{c}$ and free volume $\nu$, are determined 
by $V(r)$, they reenter into the dynamics by their influence on molecular 
mobility. Thus two factors govern dynamics, the capacity for 
intermolecular coupling directly from $V(r)$, and $S_{c}$ and $\nu$ that come 
indirectly through $V(r)$. On supercooling a liquid, $S_{c}$ and $\nu$ change, and 
since the kinetic fragility $m$ is the slope of $T_{g}$-scaled temperature 
variation of $\tau$, it is unsurprising that $m$ is correlated with the slope of the 
corresponding change of $S_{c}$, i.e., the thermodynamic fragility. The 
capacity for intermolecular coupling of $V(r)$ is solely responsible for the 
shape of the dispersion or the nonexponentiality parameter $(1-\beta)$, and it also 
determines $\tau$ in conjunction with $S_{c}$ and $\nu$. The results of our simulation 
with the three potentials support this view.  Increasing the density of the 
particles of the binary Lennard-Jones system with the fixed (12,6) potential 
effectively forces the particles to be closer to each other and thereby 
increases intermolecular coupling. The simulation performed in this manner~\cite{Sastry2001}
showing that $m$ increases with density can be 
reinterpreted as due to increase of intermolecular coupling.
Intermolecular
coupling manifests itself in the dynamic properties in various ways. The
most direct way is the width of the dispersion measured by $(1-\beta)$. The others
include~: (1) the $Q^{-2/(1-\beta)}$ dependence of $\tau$
obtained by quasielastic neutron scattering~\cite{Ngai2000},
and (2) the proportionality~\cite{Ngai2003} between $(1-\beta)$ and
$\left( \log(\tau) - \log \left( \tau_{\beta} \right) \right)$
at constant $\log(\tau)$ or at $T_{g}$ (where $\tau_{\beta}$ is the Johari-Goldstein relaxation time).
All these are evidences for intermolecular coupling 
that must be taken into consideration in conjunction with $S_{c}$ and $\nu$ for 
explaining all observed experimental facts in the liquid state. At low 
temperatures and deep in the glassy state, $S_{c}$ and $\nu$ having constant 
values cannot influence the temperature dependence of the vibrational 
properties characterized by $\alpha$. Hence, $\alpha$ is controlled by the anharmonicity of 
$V(r)$, as demonstrated by the simulations.

In summary, we demonstrate by using three different interparticle potentials 
of binary Lennard-Jones systems that the capacity for intermolecular 
coupling and anharmonicity of the potential are responsible for the 
correlations between various dynamic, thermodynamic and vibrational 
properties of glass-formers. Increase of the capacity for intermolecular 
coupling and anharmonicity has the effects of increasing the kinetic 
fragility, $m$, and the nonexponentiality parameter, $(1-\beta)$, in the liquid state, 
and of increasing in the glassy state the parameter $\alpha$
that characterize the $T_{g}$-scaled temperature dependence of the nonergodicity 
parameter determined by vibrations at low temperatures. The correlations 
between $m$, $(1-\beta)$, $\alpha$ and other quantities follow as consequences, and 
their observations by experiments explained. 

\textit{Acknowledgments.} --- The authors wish to thank for the use of the computational facilities of the IDRIS 
(Orsay, France) and the CRI (Villeneuve d'Ascq, France).
This work was supported by the INTERREG III (FEDER) 
program (Nord-Pas de Calais/Kent). Work at NRL was supported by the Office 
of Naval Research.

\end{document}